# DEALING IN PRACTICE WITH HOT-SPOTS

R. Moretón[1], E. Lorenzo[1], J. Leloux[2], J.M. Carrillo[1]
[1]Instituto de Energía Solar – Universidad Politécnica de Madrid, Photovoltaic Systems Group, EUITT, Madrid, Spain
[2]WebPV, Madrid Spain

ABSTRACT: The hot-spot phenomenon is a relatively frequent problem occurring in current photovoltaic generators. It entails both a risk for the photovoltaic module's lifetime and a decrease in its operational efficiency. Nevertheless, there is still a lack of widely accepted procedures for dealing with them in practice. This paper presents the IES-UPM observations on 200 affected modules. Visual and infrared inspection, electroluminescence, peak power and operating voltage tests have been accomplished. Hot-spot observation procedures and well defined acceptance and rejection criteria are proposed, addressing both the lifetime and the operational efficiency of the modules. The operating voltage has come out as the best parameter to control effective efficiency losses for the affected modules. This procedure is oriented to its possible application in contractual frameworks.

## 1 INTRODUCTION

A hot-spot consists of a localized overheating in a photovoltaic (PV) module. It appears when, due to some anomaly, the short circuit current of the affected cell becomes lower than the operating current of the whole, giving rise to reverse biasing, thus dissipating the power generated by other cells in the form of heat. Figure 1 shows an infrared (IR) image of one hot-spot. The anomalies that cause hot-spots can be external to the PV module: shading [1] or dust [2]; or internal: micro-cracks [3-4], defective soldering [3,5], PID [6]... In general, when a hot-spot persists over time, it entails both a risk for the PV module's lifetime and a decrease in its operational efficiency [3-4,7].

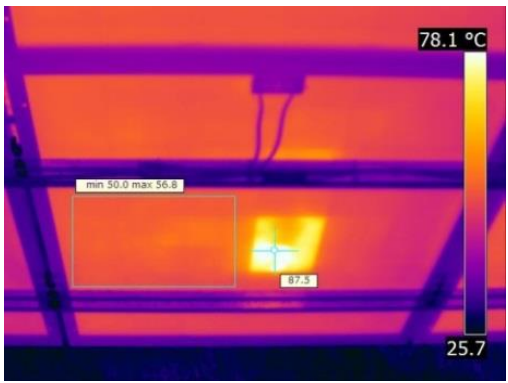

**Figure 1**. Hot-spot caused by micro-cracks. The operating temperature of the hot-spot is 87 ºC while the mean temperature of the rest of the module is 53 ºC, which represents a temperature difference of 34 ºC.

Hot-spots are relatively frequent in current PV generators and this situation will likely persist as the PV module technology is evolving to thinner wafers, which are prone to developing micro-cracks during the manipulation processes (manufacturing, transport, installation, etc.) [8]. Fortunately, they can be easily detected through IR inspection, which has become a common practice in current PV installations [4,9]. However, there is a lack of widely accepted procedures for dealing with hot-spots in practice as well as specific criteria referring to the acceptance or rejection of affected PV modules in commercial frameworks. For example, the hot-spot resistance test included in IEC-61215 is successfully passed if the module resists the hot-spot condition for a period of 5 hours, which suggests that this standard addresses transitory hot-spots, as those caused by also transitory shading, but not permanent ones, caused by internal module defects [10]. Along the same lines, the IEC 62446 only recommends to investigate the performance of all modules with significant hot-spots [11]. Furthermore, a draft of the IES-60904-12 clearly establishes how to capture, process and analyse the IR images, but still does not set out any PV module acceptance/rejection criteria [12].

This paper addresses both the lifetime and the operational efficiency of PV modules with hot-spots. Starting from the observations of 200 affected modules as experimental support, hot-spot observation procedures and well defined acceptance/rejection criteria are proposed, looking for its possible application in contractual frameworks.

## 2 FUNDAMENTALS OF HOT-SPOTS

For explanation purposes, we first consider the case of a group of $n$ identical solar cells, associated in series and protected by a by-pass diode (Figure 2-a). The operating conditions: incident irradiance, $G$, operating cell temperature, $T_C$, and polarization voltage, $V$, are such that a certain current, $I_C$, is circulating through these cells. A hot-spot appears in a cell (Figure 2-b) when some defect (micro-crack, shade, etc.) reduces its corresponding short circuit current, $I_{SC,D}$, so that

$$I_{SC,D} < I_C \qquad (1)$$

which forces the cell to operate at a negative voltage,

$$V_D = -(n-1)V_{ND} + V \qquad (2)$$

where subscripts "$_D$" and "$_{ND}$" refer, respectively, to defective and non-defective cells. Consequent power dissipation heats the defective cell, giving rise to a hot-spot, characterized by the temperature increase of this cell in relation to the non-defective ones, $\Delta T_{HS}$. The by-pass diode assures $V \geq 0$, thus limiting the negative biasing and the power dissipation in this cell. Obviously, the maximum hot-spot temperature is attained when the group is short-circuited or, which is nearly the same, when the bypass-diode is ON. Note that $\Delta T_{HS}$ is directly related to the product $I_C \times V_D$. In other words, hot-spot temperature mainly depends on the operating voltage and incident irradiance (which modulates $I_C$), on the defect gravity (which determines $I_{SC,D}$) and on the second quadrant *I-V* characteristic of the defective cell (which modulates $V_D$). As this characteristic can substantially differ from one cell to another, even within the same PV module, the hot-spot temperature also depends on the particular defective cell [13].

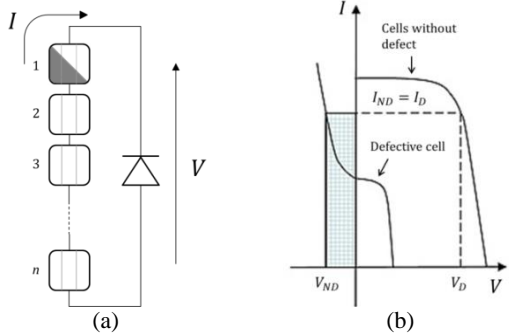

**Figure 2.** (a) Electrical connection of *n* originally identical cells protected by a by-pass diode. One of the cells is affected by a shade or an internal defect that limits its short-circuit current. (b) I-V curve of both the affected cell and the non-affected ones.

Now, let us consider the case of a PV module made up of three series associated groups, each made up of *n* cells and a bypass diode (Figure 3-a). Note that many currently commercial PV modules respond to this configuration, with *n* ranging typically from 20 to 24. A defective cell like the one described above does not reduce now the PV module sort-circuit current but becomes an anomalous step in the first quadrant of the *I-V* and *P-V* curves (Figure 3-b).

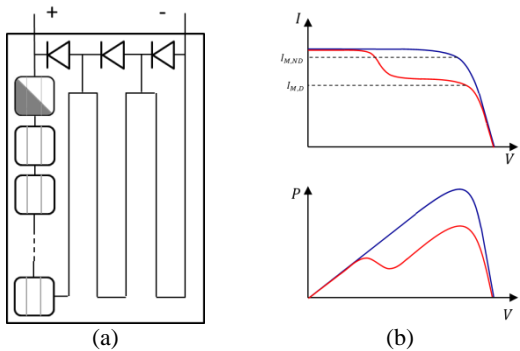

**Figure 3**. (a)Electrical scheme of a PV module with 3 groups, each made up of *n* cells and a by-pass diode. (b) *I-V* and *P-V* curves of a defective and a non-defective module. Observe the difference in the current at the maximum power point.

Again, $\Delta T_{HS}$ depends on the operating voltage of the concerned group, which, in turn, depends on the operating voltage of the PV module. The voltage at the step marks the bypass diode turning ON, and $\Delta T_{HS}$ reaches its maximum for the voltage range below this step. Figure 4 shows examples of *I-V* curves of real modules affected by hot-spots. It is worth noting that current at the maximum power point of the defective module, $I_{M,D}$, is always lower than that corresponding to the non-defective ones, $I_{M,ND}$:

$$I_{M,D} < I_{M,ND} \qquad (3)$$

Furthermore, if a module like these is connected in series with many other modules (often between 20 and 30 modules) and the resulting string is connected to an inverter able to impose the Maximum Power Point (MPP), the operating current of the group must range from between $I_{M,ND}$ and $I_{M,D}$. Then, the larger the number of modules in the series, the closer the operating current will be to $I_{M,ND}$. In this situation, the operating voltage of the defective module is well below that corresponding to its MPP. The important thing to remember is that the power loss of a defective PV module is much larger when it works associated to other non-defective modules than when it works alone. A practical consequence of the latter is that this module could pass the standard warranty conditions (referring to the maximum power of the module alone) while failing to deliver the power in practice.

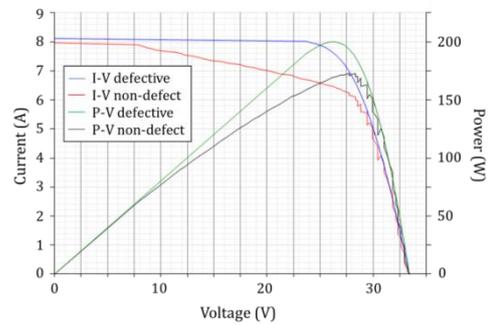

**Figure 4**. I-V curve of a defective module affected by a fill-factor loss.

Finally, not only defective cells but also defective by-pass diodes can bring about hot-spots. In the latter case, short-circuited diodes give rise to an easily recognizable thermal pattern, consisting of an anomalous hotter band, somewhat like a brushstroke extended over the cells protected by the affected diode, with several cells exhibiting temperature differences of about 5 °C. Figure 5 shows an example of a PV module with a conducting by-pass diode.

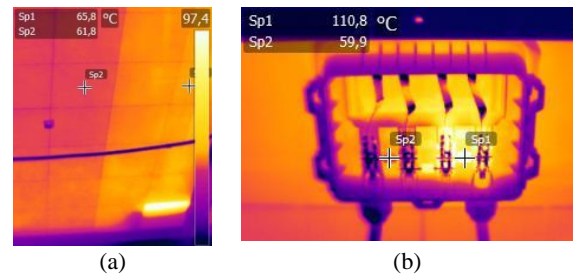

**Figure 5**. (a) PV module with one conducting by-pass diode. The cells protected by the diode are 4 °C hotter than the rest of the cells. (b) Close view of the connection box. The affected diode is at 110 °C while the others are working at 70 °C.

This is because the solar cells that make up real PV modules are not completely identical, but have a certain electrical characteristic mismatch that becomes a dispersion of voltage. At the short-circuit condition imposed by the defective diode, the sum of the voltage of all the cells protected by it is null, leading some cells becoming positive biased and others becoming negative biased. In this situation, the latter are slightly hotter than the former. Obviously, despite the temperature difference remaining low, such a module loses effective power, at a ratio equal to the number of defective diodes divided by the total number of diodes.

Because of the aforementioned dependence on $\Delta T_{HS}$ with irradiance, it is appropriate to characterize hot-spots through a value normalized to the standard irradiance, $G^* = 1000$ W/m$^2$.

$$\Delta T_{HS}^* = \Delta T_{HS} \times G^*/G$$

where $^*$ stands for the Standard Test Conditions (STC). Up to now, there has not been a widely accepted correlation for considering this effect on the heating of PV modules [12]. Nevertheless, we think that there is a certain advantage of assuming that the temperature difference is proportional to the incident irradiance. Non-linearities in the $\Delta T_{HS} - G$ relationship are likely to be small for the relatively narrow irradiance range defined by $G > 700 \, W/m^2$, which is the condition that we have imposed on our IR images.

Finally, it should be mentioned that slight temperature differences also appear in non-defective modules, mainly due to differences in heat dissipation. For example, the cells near the frame tend to be cooler while the cells around the connection box tend to be hotter. In our case, we propose $\Delta T_{HS}^* = 10\ °C$ (4 °C due to the variation in the cell efficiency in the first quadrant and 6 °C due to dissipation differences) as a minimum threshold to consider the PV module as possibly defective.

## 3 EXPERIMENTAL OBSERVATIONS

In this work, we have analysed a sample of 200 defective PV modules from two PV plants located at Cuenca and Cáceres (Spain), respectively, 122 poly-crystalline silicon modules from one single manufacturer (p-Si1) and 78 mono and poly-crystalline silicon modules from two manufacturers (m-Si and p-Si2). These defective modules were selected on the basis of a previous IR report made by the maintenance personnel of the PV plants. Then, we carried out the following tests: visual inspection, IR inspection, electroluminescence (EL), peak power and operating voltage. The Cuenca PV plant (12 MW) has been in operation since September 2011. Hot-spots soon appeared, but the module manufacturer agreed to substitute all the modules exhibiting $\Delta T_{HS}^* > 30\ °C$ on March 2013. The IR inspection that led to the selecting of the sample of defective modules was carried out on June 2013 and the tests performed by IES-UPM on January 2014. The process was similar for the Cáceres PV plant (8 MW). The operation start-up was in September 2008, the modules with hot-spots larger than 30 °C were substituted on June 2010, the IR inspection leading to the detection of the 78 defective modules took place in July 2012 and, finally, the IES-UPM tests were carried out in May 2013. It is worth noting that, in the case of the Cuenca PV plant, the initial IR inspection was made in the summer while the tests were carried out the following winter, while in the case of the Cáceres PV plant both inspections took place near the summer months. We will later discuss the consequences of these differences.

3.1 Visual inspection

Figure 6 show examples of visible defects, where micro-cracks cause a current drift and a corresponding heat that leads to the burning of the metallization fingers and in bubbles at the rear of the modules. However, we found observable defects in only a 19% of the concerned PV modules, which is too weak a correlation for considering visual defects as a basis for dealing with hot-spots.

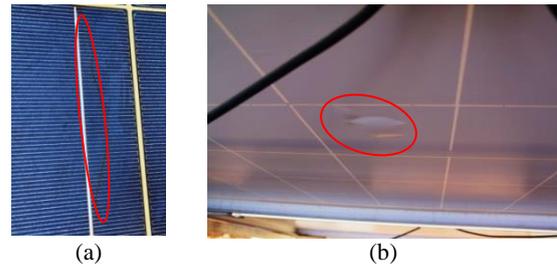

**Figure 6**. (a) Burnt metallization fingers caused by micro-cracks (b) Bubbles at the rear part of a PV module affected by hot-spots.

3.2 Infrared inspection

We obtained the IR images by means of an infrared camera (FLIR-E60). As the relevant parameter in this test is more the temperature difference than the absolute temperature value, imaging can be done at either the front or the back of the module. Just for convenience, we did all of them at the rear. Figure 7 shows the frequency distribution of $\Delta T_{HS}$. This does not reflect the total hot-spot occurrence, but only the hot-spots observed some months after the substitution of all the modules with $\Delta T_{HS}^* > 30\ °C$. Hence, the distribution tail beyond this value is a clear symptom of hot-spot time evolution. We did not observe any PID phenomena (which typically lead to a recognizable spatial pattern), thus most hot-spots are likely to be due to micro-cracks and depend on the temperature of the module, as the thermal stress affects the contact resistance between the two sides of the crack. Hence, an evolution of $\Delta T_{HS}^*$ is to be expected over the year, being larger in summer than in winter. On the other hand, daily thermal cycling typically entails degeneration, leading to a probable worsening of hot-spots over time. However, these are not absolute rules. Each micro-crack is somewhat unique and even an improvement with thermal cycling can be observed [8].

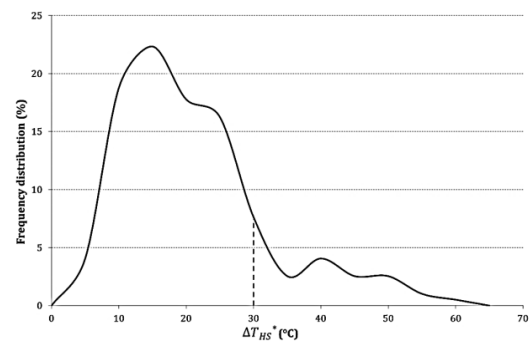

**Figure 7**. Frequency distribution of the temperature difference in the PV modules with hot-spots. The values with $\Delta T_{HS}^* > 30\ °C$ reflect the hot-spot evolution.

Figure 8 shows the combined result of these effects. Each point in the graph describes the observed $\Delta T_{HS}^*$ at two different moments. Figure 8(a) shows the evolution at the Cáceres PV plant between July 2012 (average ambient temperature, $T_A = 34\ °C$) and May 2013 ($T_A = 25\ °C$). All the modules showing $\Delta T_{HS}^* > 5\ °C$ in July have been considered. Despite the dispersion being high, on average, $\Delta T_{HS}^*$ has increased 11%. Figure 8(b) shows the case at the Cuenca PV plant between June 2013 ($T_A = 28\ °C$) and January 2014 ($T_A = 10\ °C$). Only those modules with $\Delta T_{HS}^* > 15\ °C$ in June have been considered on this occasion. Here, the average $\Delta T_{HS}^*$ has decreased by 22%, in an example of seasonal effects overcoming the degradation over time.

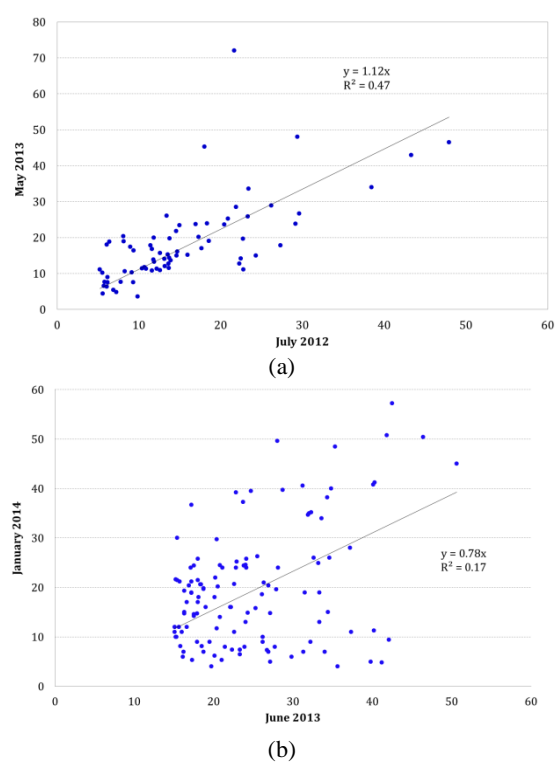

(a)

(b)

**Figure 8**. Hot-spot temperature evolution. Each point corresponds to $\Delta T_{HS}^*$ of a particular module at two different moments. At the Cáceres PV plant (a), both took place during hot months. A general $\Delta T_{HS}^*$ increase over time is noticeable. At the Cuenca PV plant (b), the latter moment corresponds to a colder month than the former. In this case, an average $\Delta T_{HS}^*$ decrease can be observed.

3.3 Electroluminescence

The objective of this test was to analyse the correlation between the portion of isolated area of a cell affected by micro-cracks and the magnitude of hot-spots. The analyses were carried out directly in the field during night using an EL camera (pco.1300 solar) and a power source. Each module was polarized in the fourth quadrant at 25% of the STC rated short circuit current. The experiment was carried out in January 2014 and applied only to a smaller sample of 35 PV modules in the Cuenca PV plant, due to the difficulties of implementing this test on site. We have followed the crack type classification proposed by Köntges et al. [8], dividing the affected cells into C-type (those exhibiting only background noise for the inactive cell part) and B-type (those exhibiting a reduced intensity but higher than the background noise). Figure 9 shows an example of an EL image obtained in the field and figure 10 shows the relationship between the fraction of cell that is isolated and the temperature difference.

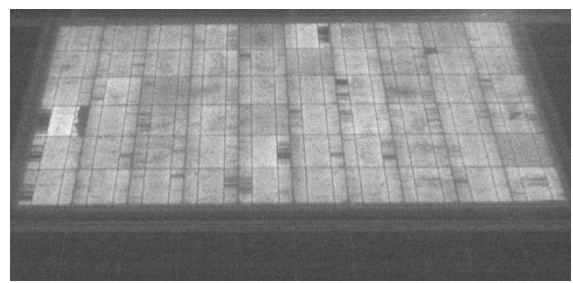

**Figure 9.** Electroluminescence image of a hot-spot affected PV module obtained in the field. Two cells with appreciable isolated areas can be observed (nearly a 40% for the left side cell – 20% B-type and 20% C-type crack – and almost 20% for the upper side cell – B type crack).

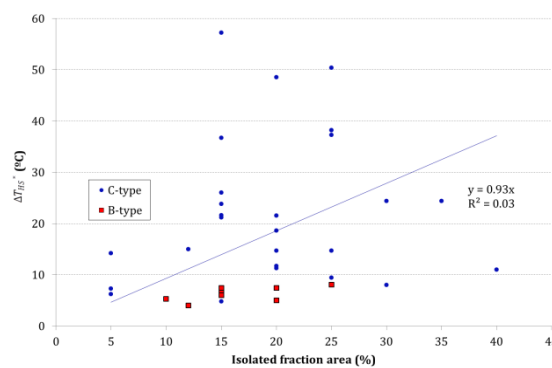

**Figure 10**. Relationship between $\Delta T_{HS}^*$ and the fraction of cell isolated by a crack. Squares and circles represent B-type and C-type cracks respectively.

We observed that all the modules showing a hot-spot in the summer IR inspections had some micro-crack in the affected cell but none of the cells with B-type cracks generated a hot-spot in winter. A proportional but very weak trend between the isolated area and $\Delta T_{HS}^*$ ($R^2=0.03$) was found. The relationship between the isolated fraction area and the power loss of the module, which remained very weak ($R^2=0.05$) was also analysed. A possible explanation is that the contact resistance between the two sides of the micro-crack varies with module temperature and can thus be much larger during the day (when hot-spots are observed) than during the night (when EL are obtained). Then, some areas can be miss-classified, leading to an incorrect estimation of the hot-spot problem. Whichever the case, EL images, despite being a very useful tool for quality control during the PV manufacturing processes, are not appealing for dealing with hot-spots in the field.

3.4 Electrical inspections: power and operating voltage

Individual I-V curves of all the affected PV modules were obtained with a commercial I-V tracer (Tritec Tri-ka) and extrapolated to STC in accordance with the IEC-60891 (procedure 1), using the current and voltage temperature coefficients given by the manufacturer. Around 53% of the modules presented some anomalies in the I-V curve, as steps or an abnormally low fill factor. Figure 11(a) shows the relationship between $\Delta T_{HS}^*$ and the power loss in respect to the manufacturer's flash value, for 50 PV modules of the Cuenca PV plant. The high spread can be observed as can the fact that most of the modules satisfied the usual power warranty condition (typically, 90% of the minimal rated power output after 10 years). However, this is scarcely representative of their in-field behaviour, which is better appreciated through the operating voltage of the module, when working within the PV array. The latter was measured by simply inserting "T" connectors into the module output wires. Then, the voltage losses as regards the non-defective modules can be understood directly as power losses, as the current is common for all the modules connected in series. Figure 11(b) shows the relationship between the power loss and the operating voltage loss for the same 50 modules. As can be observed, the effective losses are a 55% higher than the power losses when considering the module alone.

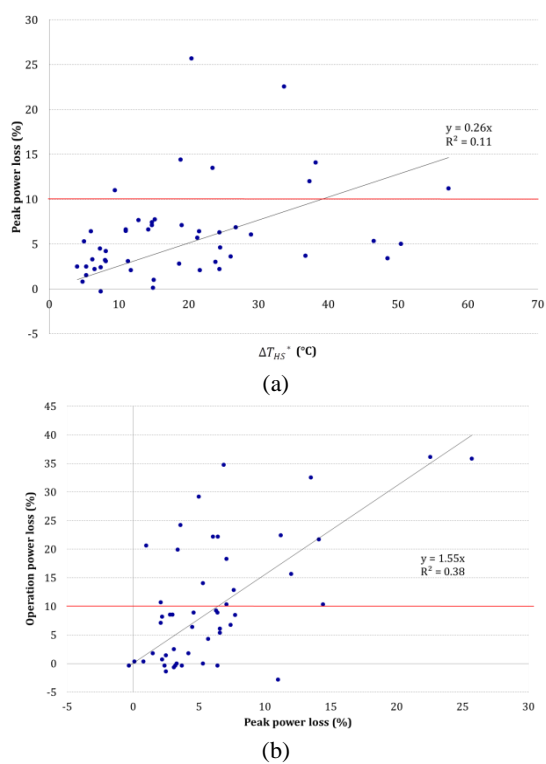

**Figure 11**. (a) Relationship between the temperature difference and the power loss for 50 PV modules. 8 of them are out of warranty conditions. (b) Relationship between the power loss and the operating voltage loss (effective power loss). In this case, 19 modules do not comply with warranty requirements.

Two key observations can be outlined. First, the standard peak power is not a good indicator of the energy production capacity of defective modules, so that it must be disregarded for dealing with hot-spots. Second, the correlation between $\Delta T_{HS}^*$ and $\Delta V_{HS}^*$ and thus, power losses during operation, is positive, but the large dispersion does not allow the correlation at individual levels to be applied. In other words, the power loss of a defective module must be deduced from direct voltage measurements not from thermal observations. Apart of that, figure 12 shows the relationship between the temperature difference and the operating voltage loss for a more complete ensemble of the 113 PV modules of the three different manufacturers.

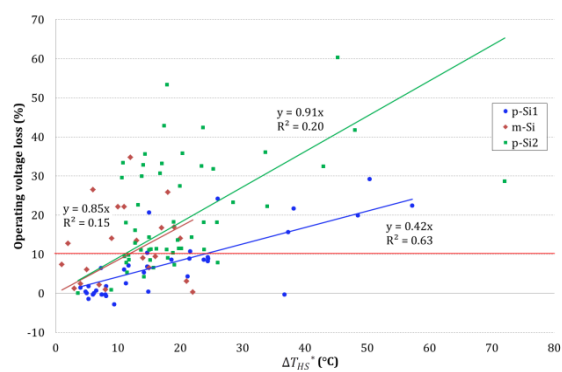

**Figure 12**. Relationship between the temperature difference and the operating voltage loss for 113 modules from 3 different manufacturers.

It can be observed that the behaviour is not the same for every manufacturer (neither in the correlation slope nor in the spread around it). The correlation between operating voltage loss and temperature difference is stronger in the case of module p-Si1 ($R^2=0.63$) and weaker for the cases of modules m-Si and p-Si2. These divergences likely reflect differences in the original material as well as non-uniform degradation affection due to different operation times (3 years in the case of module p-Si1 and 5 years for modules m-Si and p-Si2). Whichever the case, this behaviour spread is not relevant here.

4 DISCUSSION

Hot-spots threaten the PV module lifetime, as degradation processes are generally accelerated by temperature. In particular, encapsulate discoloration and browning, and delamination [14]. Previous experiences do not allow a clear relation between module temperature and lifetime [7] to be established. Therefore, in order to set a maximum acceptable value, $\Delta T_{HS,MAX}^*$, we must rely on intuitive but reasonable approaches. We propose to consider 85°C, which is the maximum temperature of the thermal cycling tests described in the IEC-61215 as the maximum absolute PV module temperature for acceptance/rejection purposes. This limit has been also proposed by other authors [7]. Then, $\Delta T_{HS,MAX}^*$ should be thus so as to guarantee that the hot-spot absolute temperature remains below that limit. Figure 13 shows the annual frequency distribution of the day-time operating temperature in the Cuenca PV plant, which can be considered as representative of a Mediterranean climate (characteristic of Southern Europe and some parts of USA, Australia or South America). The maximum cell temperature is 70 °C and the 99-percentile temperature is 65 °C. As these high temperatures are also associated to high irradiances, setting $\Delta T_{HS,MAX}^* = 20\ °C$ limits the time above 85 °C to around 40 hours a year (1% of the time) for these climate conditions, which seems a reasonable commitment. Moreover, it avoids reaching 100 °C, which has been sometimes suggested as an absolute maximum for preventing early degradation [15].

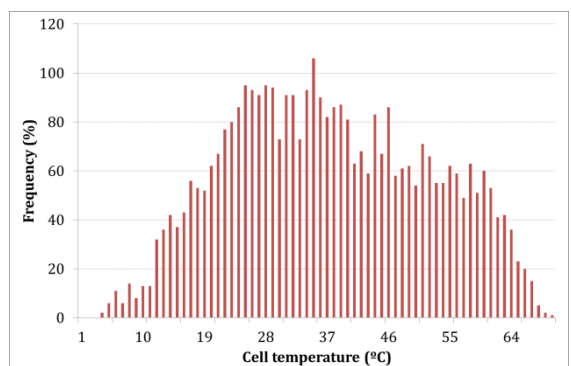

**Figure 13**. Annual frequency distribution of the operating temperature at the Cuenca PV plant

As regards energy losses, it seems logical to just extend the application of usual warranties to defective modules. Hence, it is proposed to reject any module exhibiting hot-spots whose corresponding voltage losses (in relation to a non-defective module being part of the same series association) within the PV system in normal operation, exceeds the allowable peak power losses fixed at standard warranties. This is also applicable to PV modules with defective by-pass diodes, regardless the temperature of the derived hot-spot.

## 5 CONCLUSIONS

There is still not a widely accepted reference on how to face the hot-spot problem within commercial frameworks. Supported by experimental observations on 200 PV modules exhibiting hot-spots, the following procedure is proposed as a practical in-field approach to accomplish IR imaging inspection:
1) Assure $G > 700$ W/m$^2$
2) Perform the analyses in summer, preferably on the hottest days
3) Extrapolate the temperature difference, $\Delta T_{HS}^*$, considering a lineal relationship with the irradiance.

Then, for every PV module with a hot-spot, the following is proposed:
1) If $\Delta T_{HS}^* < 10°C$, to consider the module non-defective, except in the case that one or more by-pass diodes are defective.
2) If $\Delta T_{HS}^* > 20°C$, to consider the module defective.
3) If $10°C < \Delta T_{HS}^* < 20°C$, to consider all the modules with an effective power loss (measured as a decrease in the operating voltage in relation to a non-defective module of the same string) that exceeds the allowable peak power losses fixed at standard warranties defective.

Finally, it is worth mentioning that this procedure and acceptance/rejection criteria have already been applied by IES-UPM when mediating in hot-spot conflicts between module manufacturers and engineering, procurement and construction companies during the last years.